\documentstyle[aps,twocolumn,epsfig]{revtex}
\begin{document}

\title{Topology dependent quantities at
the Anderson transition}

\author{Keith Slevin}
\address{Dept. of Physics, Graduate School of Science,
Osaka University, \\ 1-1 Machikaneyama, Toyonaka,
Osaka 560-0043, Japan}

\author{Tomi Ohtsuki}
\address{Department of Physics, Sophia University,
Kioi-cho 7-1, Chiyoda-ku, Tokyo 102-8554, Japan}

\author{Tohru Kawarabayashi}
\address{Department of Physics, Toho University,
Miyama 2-2-1, Funabashi 274-8510, Japan}

\maketitle

\abstract{The boundary condition dependence of the critical
behavior for the three dimensional Anderson 
transition is investigated.
A strong dependence of the scaling function and the
critical conductance distribution on the boundary
conditions is found, while the critical disorder and
critical exponent are found to be independent of
the boundary conditions.}


The Anderson transition is a continuous quantum phase 
transition which separates metallic and insulating phases of the 
non- interacting electron gas \cite{ANDERSON,LR}.
It is expected that the critical behavior should depend 
only on the basic symmetry of the Hamiltonian under the 
operation of time reversal \cite{WEGNER,GANG}.
Recently this was clearly confirmed by numerical 
simulation \cite{SO}.

At the transition the correlation length diverges and  
quantities such as level statistics \cite{SSSLS,ZK}
and the conductance distribution \cite{SO,SHAPIRO,MARKOS,MARKOS2}
become size independent and universal.
The discovery that the critical level spacing distribution 
\cite{BRAUN,YUDSON}, and indications that the critical
conductance distribution \cite{SOUKOULIS,SO2} also, depend
on the boundary conditions was unexpected.

In this letter, we analyze both the corrections to scaling
induced by different boundary conditions and the effect
of the different boundary conditions on the critical disorder,
critical exponent and critical conductance distribution.
None of the boundary conditions we consider break time 
reversal symmetry, so no change in the critical behavior can 
be predicted on the general grounds of a transition between 
universality classes. 

We report the results of two different simulations. In the first
the scaling behavior of the localization length of electrons
on long quasi-$1D$ bars is examined.
In the second the conductance distributions for ensembles of
cubes of disordered material in a two probe measuring geometry
are determined. 
Both simulations were repeated for three different
boundary conditions;
pbc) periodic boundary conditions in both transverse directions,
mbc) periodic boundary conditions in one direction and fixed 
boundary conditions in the other and
fbc) fixed boundary conditions in both transverse directions.
These boundary conditions can be thought of as corresponding to
different topologies; fbc corresponds to the topology of a wire,
while mbc corresponds to that of a hollow cylinder.

We find that the location of the mobility edge separating
the localized and diffusive phases is unaffected by the choice of
boundary condition. This is also true for the critical
exponent. However, the scaling function of the localization
length and the critical conductance distribution are found
to depend strongly on the choice of boundary condition and hence
on the topology of the sample.

For the numerical simulations we have used the Anderson model
\begin{equation}
 H = V \sum_{<i,j>} C_i^{\dagger}C_j +
     \sum_i W_i C_i^{\dagger}C_i ,
\end{equation}
where $C_i^{\dagger}(C_i)$ denotes the creation (annihilation)
operator of an electron at the site $i$ of a 3D cubic lattice.
Energies $W_i$ denote the random potential distributed 
independently and uniformly in the range
$[-W/2, W/2]$.
The hopping is restricted to nearest neighbors and its
amplitude is assumed to be the energy unit, $V=1$.

We consider first a quasi-$1d$ system, i.e., a long bar, with cross section
$L \times L$.
The standard transfer matrix technique allows
us to calculate the localization length of electrons
$\lambda(E_F,W,L)$ on the bar within a desired accuracy \cite{MK,KM}. 
The dependence of the quantity
\begin{equation}
\Lambda=\frac{\lambda(E_F,W,L)}{L}
\end{equation}
on the width of the bar is then analyzed using the finite size 
scaling method.
In this simulation we set the Fermi energy  
at the band center $E_F=0$ and vary the strength of the random potential $W$ and
the cross section size $L$. 

\begin{figure}
\epsfig{file=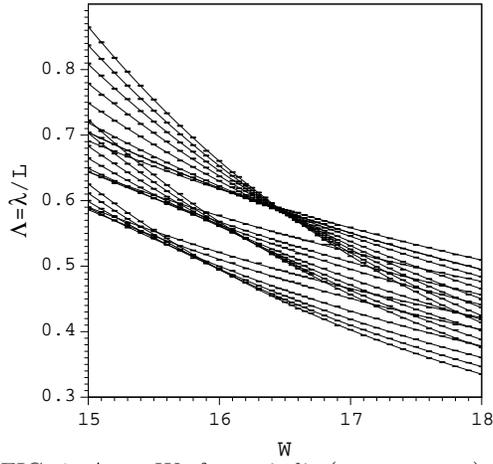,width=6.5cm}
\caption{$\Lambda$ vs. $W$. for periodic (upper curves),
mixed (middle curves) and fixed (lower curves) boundary
conditions.}
\label{F1}
\end{figure}

The data obtained are shown in Fig.\ref{F1}.
In the absence of any corrections to scaling, plotting $\Lambda$ vs $W$ 
should show the critical disorder $W_c$ as the common crossing 
point of the data.
However, as seen in Fig.\ref{F1}, the curves for different sizes
do not cross at a common point.
For pbc a previous analysis \cite{SO3} has suggested that main 
reason for this is the existence of a correction due to an 
irrelevant scaling variable.
For mbc and fbc surface and edge contributions might also be important.
Such corrections are irrelevant. For example, for a surface effect
we expect the corrections vanish as $L^{-1}$.

To take account of corrections to scaling we use the 
method described in \cite{SO3}.
The data are fitted to the scaling form
\begin{equation}
\Lambda=F(\psi L^{1/\nu},\phi L^{y})
\label{eq_renorm}
\end{equation}
where $\nu$ is the critical exponent describing the
divergence of the localization length,
$\phi$ is the leading irrelevant variable and $y$ its
irrelevant exponent.
The best fit is found by minimizing the $\chi^2$ statistic
in the usual way.

When fitting the data, Eq.(\ref{eq_renorm}) is expanded in a 
Taylor series to first order in the irrelevant variable.
\begin{equation}
\Lambda=F_0(\psi L^{1/\nu})+ \phi L^y F_1(\psi L^{1/\nu}) 
\label{eq_expansion}
\end{equation}
Both $F_0$ and $F_1$ are expanded to third order in their
arguments. 
The relevant and irrelevant scaling variables are expanded 
in power series of the dimensionless disorder
$w=(W_c-W)/W_c$ as follows  
\begin{equation}
\psi = \psi_1 w  \ \ \ , \ \ \  \phi = \phi_0
\label{varexpand}
\end{equation}
For pbc the expansion of the relevant field in
(\ref{varexpand}) was continued to quadratic
order as this gave a better quality of fit.
Also, for fbc, data for $L=4$ had to be omitted in order to
obtain an acceptable fit.

\begin{figure}
\epsfig{file=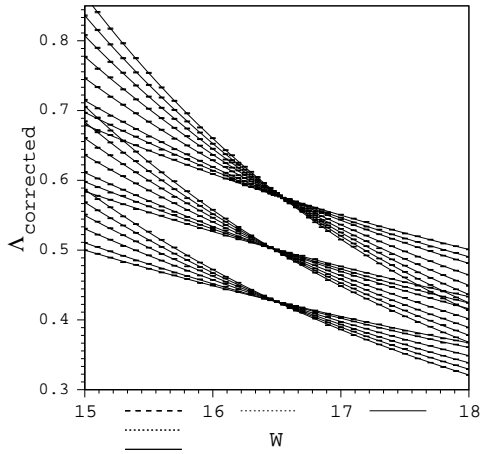,width=6.5cm}
\caption{$\Lambda$ vs. $W$ after the surface corrections
are removed.}
\label{F2}
\end{figure}

\begin{figure}
\epsfig{file=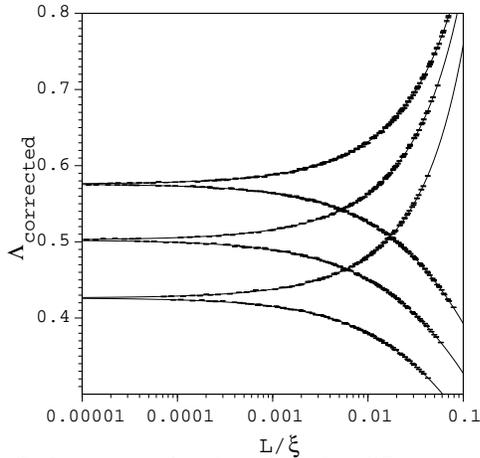,width=6.5cm}
\caption{The scaling functions for different boundary
conditions.}
\label{F3}
\end{figure}

The numerical data and the associated fits are shown in 
Figs. \ref{F1},\ref{F2} and \ref{F3}.
The estimates for the critical parameters are listed in
Table \ref{T1} and further details of the fits in Table \ref{T2}.
It can be seen from Fig. \ref{F1} that the main effect of 
imposing a fixed boundary condition at a given disorder 
is a decrease of the localization length $\lambda$.
Also the deviations from scaling behavior are much larger
than those for periodic boundary conditions. For fbc
the corrections are around 15\% for the smallest system
size, while they are only around 2\% for pbc.
The estimate of the irrelevant exponent for pbc is consistent
with that in \cite{SO3}. The estimates for mbc and pbc are both much
larger and close to $y=-1$ suggesting that the dominant correction
is a surface effect.

In Fig. \ref{F2} we see that taking account
of corrections to scaling by plotting
$\Lambda_{\mathrm{corrected}}=\Lambda - \phi L^y F_1(\psi L^{1/\nu})$
as the ordinate restores a common crossing point to
the data for different system sizes for each boundary condition.
Further the critical disorder seems to be the same for
all three boundary conditions. This is reinforced by looking
at Table \ref{T1} where $95\%$ confidence intervals are also
given.
Thus the location of the mobility edge does not seem to
be affected by the choice of boundary conditions.
The same is true for the critical exponent.

However, the scaling function, and in particular $\Lambda_c=F_0(0)$, 
do depend strongly on the boundary conditions.
The estimates of $\Lambda_c$ given in Table \ref{T1} are widely
separated with no overlap of the confidence intervals.
The change in the estimated $\Lambda_c$ is roughly 25\%, while
the analysis suggests that $\Lambda_c$ has been estimated to
within an accuracy of about 5\% for fbc and to less than 1\% for pbc.
(Note that the apparent differences in the scaling function in the
localized regime in Fig. \ref{F3} are not important; these are an artifact
of the scaling procedure in which the absolute value of the 
correlation and localization lengths are not determined.)

Next we look at the critical conductance distribution $p_c(g)$.
Since the correlation length diverges at the critical point, 
the critical conductance distribution of a phase coherent 
conductor should be scale invariant \cite{SHAPIRO}.
This was confirmed in numerical simulations \cite{SO,MARKOS} and
it was also confirmed that $p_c(g)$ depends on the universality 
class \cite{SO}.
The conductance of a classical conductor depends on the aspect ratio
(i.e. the ratio of the cross section to the length) but not
on the shape of its cross section or on its topology (i.e. whether 
it is a bar or a cylinder.)
Thus, a dependence of the critical conductance distribution of a 
quantum conductor on its aspect ratio is also expected.
Whether or not the the critical conductance distribution 
should depend on the conductor's cross sectional shape and its topology
(boundary conditions) is less clear. 

We simulated the conductance distribution for an ensemble 
of $L\times L\times L$ cubic samples in a two probe measuring geometry
using a Green's function iteration technique \cite{ANDO}.
The conductance in units of $e^2/h$ is
\begin{equation}
g = 2 \mathrm{tr}(tt^{\dagger})
\end{equation}
where $t$ is the transmission matrix found in the Green's function
iteration and the factor of two takes account of spin degeneracy.

\begin{figure}
\epsfig{file=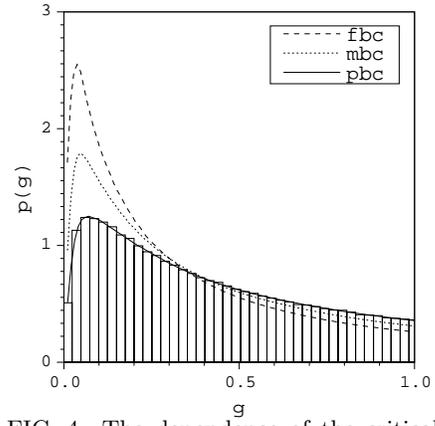,height=5.5cm}
\caption{The dependence of the critical conductance distribution on the
choice of boundary conditions.
Here the Fermi energy $E_F=0$,the system size $L=10$ and 
the disorder $W=16.54$ independent of the choice of boundary 
conditions.}
\label{F4}
\end{figure}

We first simulated an ensembles of 1,000,000 systems for each 
boundary condition.
The resulting distributions are shown in Fig. \ref{F4}.
The choice of boundary condition affects $p_c(g)$
especially for small $g$. There is a tendency towards more
insulating behavior for fbc.

\begin{figure}
\epsfig{file=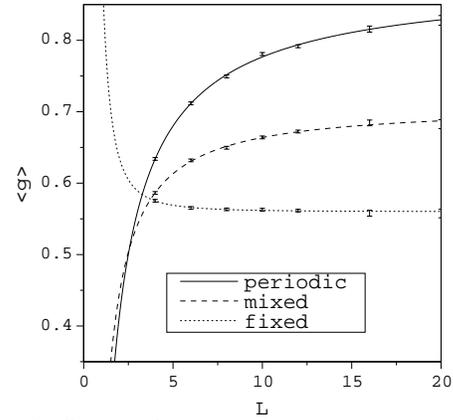,height=5.5cm}
\caption{The size dependence of $<g>$ for different
boundary conditions.
Here the Fermi energy $E_F=0.5$ for which we estimated 
$W_c=16.53$ independent of the choice of boundary conditions.}
\label{F5}
\end{figure}

It is expected on general grounds that $p_c(g)$ will be size 
independent for ``large enough'' system sizes.
This leaves open the possibility that the dependence of $p_c(g)$
on the boundary conditions in Fig. \ref{F4} may simply be an 
indication that the system sizes employed in our work are not
``large enough''.
We therefore decided to examine the size dependence of $p_c(g)$.
The dependence of $<g>$ on the size $L$ for $L=4$ to $L=20$ for each
boundary condition is shown in Fig. \ref{F5}.
Ensembles of $100,000$ systems were generated for each system
size with the exceptions of $L=16$ and $L=20$ where the ensemble
sizes were reduced to $25,000$ and $10,000$ respectively.

To estimate the asymptotic value of $<g>$ in the limit 
$L\rightarrow \infty$ we have assumed that
the size dependence is due to an irrelevant scaling variable and
fitted the data to 
\begin{equation}
<g(L)> = <g(\infty)> + a L^{y^{\prime}}.
\end{equation}
The details of the fits are given in Table \ref{T3}.
From the results it seems clear that 
the boundary condition dependence will not 
disappear as $L\rightarrow \infty$.
The asymptotic value of the mean conductance is reduced by
about 40\% while the analysis suggests that it has been
estimated to within a few percent.
The numerical data are consistent with a size independent and universal
$p_c(g)$ but also one that depends on the boundary conditions
even in the limit $L\rightarrow \infty$.

The size dependence of the critical conductance fluctuations were 
analyzed in a similar way and the results are also given in Table \ref{T3}.
(For fbc no size dependence of the variance was detected within the 
accuracy of the simulation so the value given is simply a weighted
average over different system sizes.)
The values given can be compared with the value of
var($g$)=$1.18$ (with $g$ in units of $e^2/h$) for universal
conductance fluctuations in the metallic regime \cite{LSF}.
Though the absolute magnitude of the fluctuations is smaller than
in the metallic regime, the conductance fluctuations are in fact of the
same order of magnitude as $<g>$ at the transition.

The estimates of the irrelevant exponent in Table \ref{T3}
again suggest the presence of surface effects.
However, for the conductance calculation it seems to be pbc
which presents a surface effect rather than fbc. For the latter
boundary condition a surface correction does not seem to be present
and we find a value of the irrelevant exponent which is consistent with
that for the scaling behavior of the localization length 
with pbc given in Table \ref{T1}.

In conclusion we have found that the some aspects of the
critical behavior at the Anderson transition,
the location of the mobility edge and the critical
exponent, are independent of the boundary conditions.
Other aspects, the scaling function and the critical
conductance distribution, seem to depend strongly on the
boundary conditions.
The invariance of the critical disorder is reasonable since
localized states should be unaffected by boundary conditions.
However, this argument does not hold in the critical regime
and, as we have shown, important aspects of the critical behavior 
do depend on the topology.
The exception is the critical exponent provided, as here, 
that the boundary conditions do not break time reversal symmetry.

Finally, we recall that a similar dependence of the scaling 
functions on boundary conditions has also been found in 
classical percolation \cite{HU,CARDY}. 
This suggests that the quantum nature of the Anderson transition 
might not be crucial for the existence of this sort of effect.

The authors would like to thank K.Fukushima, S.Todo, Y.Okabe and N.Hatano
for useful discussions.

\begin{table}
\noindent
\begin{tabular}{|l|l|l|l|l|}
     & $\nu$        & $W_c$        & $\Lambda_c$      &  $y$\\ \hline
pbc  & 1.56(55,58)  & 16.54(53,55) & 0.576(.574,.577) &  -2.8(3.2,2.4)\\
mbc  & 1.60(56,64)  & 16.47(42,52) & 0.502(.494,.509) &  -1.3(1.5,1.2)\\
fbc  & 1.54(41,61)  & 16.49(39,64) & 0.426(.403,.442) &  -1.2(1.4,1.0)\\ 
\end{tabular}
\caption{The best fit estimates of the critical exponent, 
the critical disorder, $\Lambda_c$ and the irrelevant exponent
together with their $95\%$ confidence intervals.}
\label{T1}
\end{table}

\begin{table}
\begin{tabular}{|l|l|l|l|l|}
 & $N_d$ & $N_p$ & $\chi^2$ & $Q$ \\ \hline
pbc      & 336 & 12 & 335 & 0.3 \\
mbc      & 238 & 11 & 233 & 0.4 \\
fbc      & 204 & 11 & 212 & 0.2 \\ 
\end{tabular}
\caption{The boundary condition,
the number of data $N_d$, the number of parameters $N_p$,
the value of $\chi^2$ for the best fit and
goodness of fit $Q$. The system sizes used were
$L=4,5,6,8,10,12,14$ with the exception of fbc where
the data for $L=4$ were omitted and pbc where data for $L=16$
are also included. The accuracy of 
the numerical data was either 0.1\% or 0.05\%.
The range of disorder used was $W=15$ to $W=18$.}
\label{T2}
\end{table}

\begin{table}
\begin{tabular}{|llll|l|}                               
 & $<g(\infty)>$ & $Q$ & $y^{\prime}$ & var$(g(\infty))$\\ \hline
pbc  & $0.89\pm .02$   & $0.2$  & $-0.9\pm.1$  & $0.472 \pm.01$\\
mbc  & $0.71\pm .01$   & $0.6$  & $-1.1\pm.1$  & $0.408\pm.003$ \\
fbc  & $0.560\pm .002$ & $0.9$  & $-2.3\pm.7$  & $0.349\pm.001$\\ 
\end{tabular} 
\caption{The estimated value of $<g>$ in the thermodynamic limit 
$L\rightarrow \infty$ together with standard errors, the goodness 
of fit $Q$ and an estimate of the irrelevant exponent.
An estimate of var($g$) in the limit $L\rightarrow \infty$ 
together with standard errors is given in the last column.} 
\label{T3}
\end{table}

\end{document}